\documentclass[conference]{IEEEtran}
\IEEEoverridecommandlockouts

\usepackage{cite}
\usepackage{amsmath,amssymb,amsfonts}
\usepackage{graphicx}
\usepackage{textcomp}
\usepackage{xcolor}
\def\BibTeX{{\rm B\kern-.05em{\sc i\kern-.025em b}\kern-.08em
    T\kern-.1667em\lower.7ex\hbox{E}\kern-.125emX}}

\usepackage{tikz,pgf}
\usepackage{amsmath,amssymb,amsfonts,bm}
\usepackage{subfigure}
\usepackage{amsthm}
\usepackage{color}
\usepackage{enumerate}
\usepackage{amsbsy}
\usepackage{amssymb}
\usepackage{amsthm}
\usepackage{amscd}
\usepackage{subfigure}
\usepackage{color}
\usepackage{cite}
\usepackage{notoccite}
\usepackage{stackrel}
\newcommand\scalemath[2]{\scalebox{#1}{\mbox{\ensuremath{\displaystyle #2}}}}

\newtheorem{rema}{Remark}

\def\bI{{\bf I}}

\begin{document}

\title{Robustness Guarantees for Optimal RIS Placement in Self-Localization Under Placement Errors
}

\author{\IEEEauthorblockN{Mahdi Koloushani}
\IEEEauthorblockA{\textit{EE Department} \\
\textit{Sharif University of Technology}\\
Tehran, Iran \\
mahdi\_koloushani@ee.sharif.edu}
\and
\IEEEauthorblockN{Omid Rezaei}
\IEEEauthorblockA{\textit{EE Department} \\
\textit{Sharif University of Technology}\\
Tehran, Iran \\
omid.rezaeihajiabadi@ee.sharif.edu}
\and
\IEEEauthorblockN{Seyed Mohammad Karbasi}
\IEEEauthorblockA{\textit{EE Department} \\
\textit{Sharif University of Technology}\\
Tehran, Iran \\
m.karbasi@sharif.edu}
\and
\hspace{100pt}\IEEEauthorblockN{Mohammad Mahdi Naghsh}
\IEEEauthorblockA{\hspace{100pt}\textit{ECE Department} \\
\hspace{100pt}\textit{Isfahan University of Technology}\\\hspace{100pt}
Isfahan, Iran \\
\hspace{100pt}mm\_naghsh@iut.ac.ir}
\and
\IEEEauthorblockN{Mohammad Mahdi Nayebi}
\IEEEauthorblockA{\textit{EE Department} \\
\textit{Sharif University of Technology}\\
Tehran, Iran \\
nayebi@sharif.edu}
}

\maketitle

\begin{abstract}
This paper studies the effect of Reconfigurable Intelligent Surface (RIS) placement errors on optimal RIS deployment for Time-of-Flight (ToF)-based self-localization. We consider a setup in which a source transmits a signal, receives the RIS-reflected echo, and estimates its own position from the corresponding ToF measurements. Under the conditions of moderate, i.i.d. Gaussian RIS position errors and i.i.d. Gaussian ToF noise, we demonstrate that the placement uncertainty manifests as a geometry-independent scaling factor in the Cramér-Rao Bound (CRB). Consequently, we prove that the optimal RIS placement remains unchanged by the presence of RIS position errors. This result is established for A-, D-, and E-optimality criteria. Simulation results verify the analysis and show that optimized RIS deployments retain their advantage even in the presence of placement uncertainty.
\end{abstract}

\section{Introduction}
Localization is a fundamental functionality in many modern wireless systems, enabling a wide range of applications including autonomous navigation, intelligent transportation, emergency response, and context-aware communications. In many emerging scenarios, it is desirable for a device to estimate its own position without relying on external anchor nodes or infrastructure. This paradigm is commonly referred to as \emph{self-localization}. Self-localization is particularly relevant in environments where the deployment of dedicated localization anchors is costly, impractical, or dynamically changing \cite{10080967 ,gezici2005localization}. 

Recent advances in Reconfigurable Intelligent Surfaces (RISs) have opened new opportunities for enhancing wireless localization performance. RIS technology enables programmable control of the propagation environment by manipulating the phase, amplitude, or polarization of electromagnetic waves using a large array of low-cost reflecting elements. By properly configuring these elements, RIS can steer, reflect, or scatter signals in desired directions, effectively shaping the wireless channel \cite{rezaei2023design, rezaei2024cooperative, wu2019towards}. RIS architectures include \emph{passive RIS}, \emph{active RIS}, \emph{Simultaneously Transmitting and Reflecting RIS (STAR-RIS)}, and \emph{Beyond-Diagonal RIS (BD-RIS)} \cite{li2025tutorial,deng2024reconfigurable}. Due to practical considerations—such as lower hardware complexity, reduced power consumption, and easier deployment—passive RIS architectures are generally more commonly adopted.

Recently, RISs have attracted significant attention in wireless localization systems due to their capability to manipulate the wireless propagation environment and improve positioning accuracy. A growing body of work has studied RIS-assisted localization frameworks in which RISs enhance conventional infrastructure-based systems by providing additional geometric diversity and improving parameter estimation performance. In this context, the work in \cite{Meng_RIS} considered a passive RIS architecture for target localization in blocked environments and investigated joint time-of-arrival and direction-of-arrival estimation while deriving the corresponding Cramér-Rao Bound (CRB) and localization algorithms. Similarly, the authors in \cite{Omar_RIS} proposed a compressed sensing framework for single-anchor localization in near-field multipath RIS-aided environments, while accounting for near-field propagation and model mismatch effects to improve localization performance. 

Beyond these infrastructure-assisted localization frameworks, recent studies have also considered RIS-enabled self-localization scenarios in which the user estimates its own position directly from RIS reflections without relying on access-points or base stations. In particular, \cite{Keyhan_self} introduced one of the first RIS-enabled self-localization frameworks with zero access-points, where the user transmits Orthogonal Frequency Division Multiplexing (OFDM) pilots and estimates its position from RIS-reflected signals while deriving the associated CRB and low-complexity estimators. More recently, RIS-enabled self-localization using Frequency Modulated Continuous Wave (FMCW) radar was experimentally demonstrated in \cite{Kim_self}, confirming the practical feasibility of RIS-based self-localization using FMCW radar measurements and RIS reflections. These studies collectively demonstrate the strong potential of RISs for enabling high-accuracy localization and self-localization in future wireless systems.

Similar to classical sensor and anchor placement problems in wireless localization systems, the geometric configuration between the source and RISs plays an important role in determining localization performance \cite{Henk_surv}. Consequently, several recent studies have investigated optimal RIS placement strategies for wireless localization systems. In particular, the authors in \cite{Bakhshi_RIS_Placemnt} studied RIS placement optimization in near-field localization scenarios by minimizing the position error bound and demonstrated that the localization accuracy strongly depends on the RIS deployment position. Similarly, the work in \cite{Mondal_RIS_Placement} investigated RIS deployment design using CRB-based optimization frameworks and analyzed the impact of RIS placement on localization accuracy. Nevertheless, existing studies generally assume ideal deployment conditions. In practice, RIS installation may be subject to positioning inaccuracies and deployment errors, causing the RISs to deviate from their intended optimal locations and consequently degrading localization performance. These considerations motivate the investigation of robust RIS placement strategies under RIS position errors.

In this work, we investigate the optimal RIS placement problem for RIS-enabled self-localization systems in the presence of RIS position errors. Specifically, we consider practical deployment scenarios in which the installed RIS locations deviate from their intended positions due to installation inaccuracies. Using a CRB framework, we analyze the impact of RIS position errors on localization performance and optimal RIS deployment. Furthermore, we prove that under mild conditions, RIS position errors result only in a geometry-independent scaling of the localization error bound. Consequently, the optimal RIS placement obtained in the absence of RIS position errors remains unchanged when RIS position errors are taken into account. This result implies that, under the considered conditions, RIS position errors can be neglected during the RIS placement optimization stage, significantly simplifying the optimal deployment design problem while preserving the optimal RIS configuration. Finally, numerical results verify the proposed analytical findings and confirm that the optimal RIS placement remains unchanged in the presence of RIS position errors under the considered conditions.

\emph{Notation:} Bold lowercase (uppercase) letters are used for vectors (matrices).
The notations $\|\cdot\|$, ${(\cdot)^{{T}}}$, $\mathrm{tr} \{ \cdot \}$, $\mathbf{\lambda}_{\mathrm{max}}(\cdot)$, $\mathrm{det}(\cdot)$, and $\mathrm{BlkDiag}(\cdot)$ denote the Euclidean norm, transpose, trace, largest eigenvalue, determinant, and block-diagonal construction, respectively. The Gaussian distribution with mean $\boldsymbol{\omega}$ and covariance $\mathbf{\Sigma}$ is denoted by $\mathcal{N}(\boldsymbol{\omega},\mathbf{\Sigma})$, while $\mathcal{U}\{a, ..., b\}$ denotes the discrete uniform distribution over the set of integers $ \lbrace a, a+1, ..., b \rbrace$. The set of ${N \times N}$ real matrices is denoted by $\mathbb{R}^{N \times N}$, and $\mathbb{R}^{N}$ denotes the set of ${N \times 1}$ real vectors. The ${N \times N}$ identity matrix is denoted by $\bI_{N}$.
\section{Signal Model}
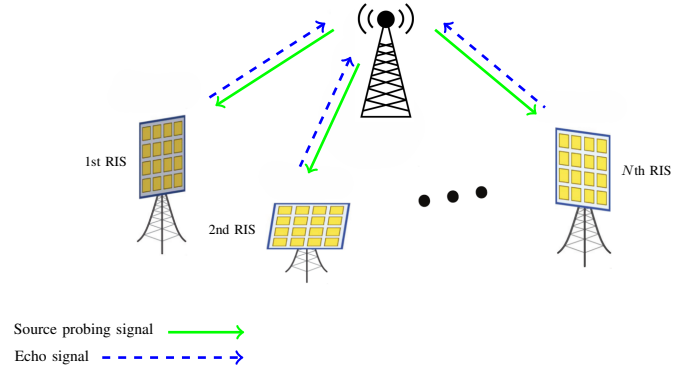
\begin{figure}
	\centering
	\begin{tikzpicture}[even odd rule,rounded corners=2pt,x=12pt,y=12pt,scale=.53,every node/.style={scale=.65}]

		\node[inner sep=0pt] (russell) at (-10,3)
		{\includegraphics[width=.6\textwidth]{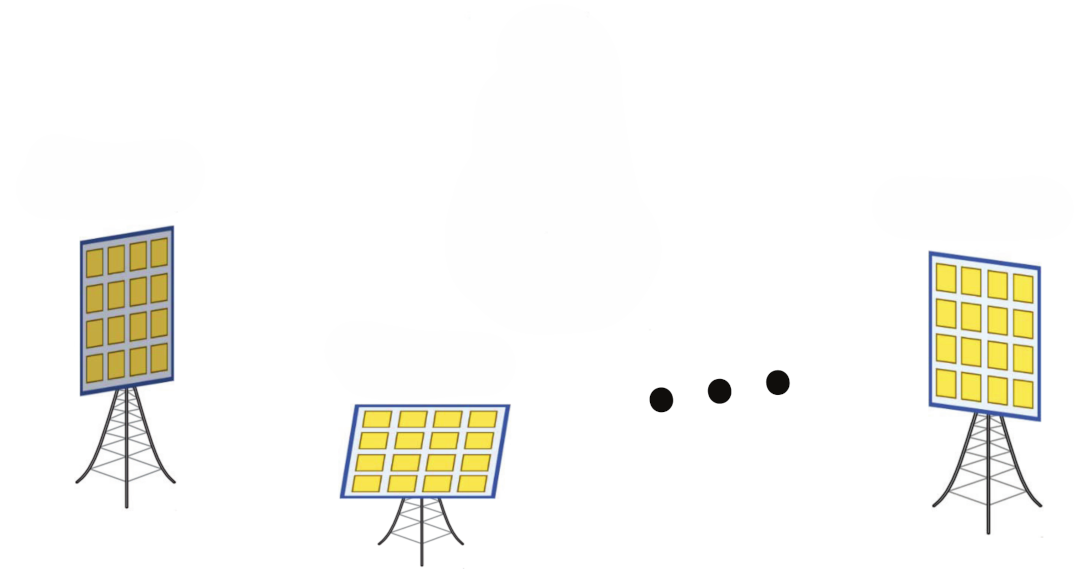}};

		\node[inner sep=0pt] (russell) at (-9,8)
		{\includegraphics[width=.065\textwidth]{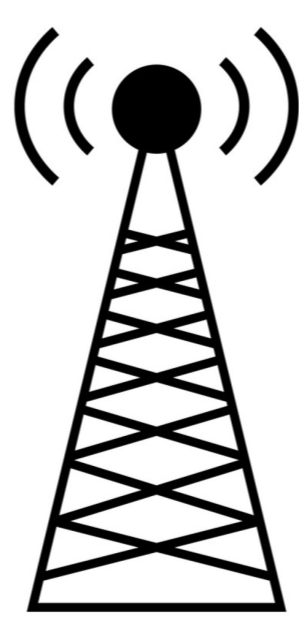}};
		\node [] at (-25.5,2) {\footnotesize 1st RIS};
		\node [] at (-18,-2) {\footnotesize 2nd RIS};
		\node [] at (6.5,1.5) {\footnotesize $N$th RIS};

		\draw[->,line width=1,green!100] (-12,9.8)--+(-7,-4.5);
		\draw[<-,dashed,line width=1,blue!100] (-12.5,10.2)--+(-7,-4.5);

		\draw[->,line width=1,green!100] (-6,9.8)--+(6,-5);
		\draw[<-,dashed,line width=1,blue!100] (-5.5,10.2)--+(6,-5);

		\draw[->,line width=1,green!100] (-10.5,7.8)--+(-3,-6.5);
		\draw[<-,dashed,line width=1,blue!100] (-11,8.2)--+(-3,-6.5);

		\draw[->,line width=1,green!100] (-14-12.1+4.3,-12.6-7+12-.5)--+(4.5,0);

		\draw[->,dashed,line width=1,blue!100] (-18.2-11.5+4,-14.1-7+12-.5)--+(8.35,0);
		
		\node [] at (-18-12.8+4,-12.5-7+12-.5) {\small Source probing signal};
		
		\node [] at (-21-11.68+4,-14.1-7+12-.5) {\small Echo signal};
	\end{tikzpicture}
	\caption{RIS-aided ToF-based self-localization.}
	\label{ht}
	\centering
\end{figure}
We consider a $D$-dimensional environment ($D=2$ or $3$), as illustrated in Fig.~1, in which a source node aims to estimate its own position using reflections from multiple passive RISs. The source transmits a probing signal and receives the echoes reflected by the RIS panels, enabling self-localization through round-trip Time-of-Flight (ToF) measurements. However, real-world deployments often introduce uncertainties in RIS positions due to factors like installation inaccuracies, environmental drifts, or calibration errors, which can substantially degrade localization performance \cite{emenonye2023ris, Zheng_geometry}. Let $\mathbf{p}_i \in \mathbb{R}^{D}$, for $i = 1, ..., N$, represent the actual positions of $N$ RIS panels. Then, the vector of all RIS positions can be denoted as $\mathbf{p}= [\mathbf{p}_1^T , \mathbf{p}_2^T, ..., \mathbf{p}_N^T]^T \in \mathbb{R}^{DN}$ which is not precisely known and follow a known Gaussian distribution $\mathbf{p} \sim \mathcal{N} (\bar{\mathbf{p}}, \mathbf{Q}_p) $, where $\bar{\mathbf{p}}$ denotes the intended (nominal) RIS positions, and $\mathbf{Q}_p$ is the covariance matrix capturing placement errors. The deterministic unknown source location is denoted by the vector $\mathbf{u} \in \mathbb{R}^{D}$. The ideal round-trip propagation delay associated with RIS $i$ is
\begin{equation}
	h_i(\mathbf{u},\mathbf{p}_i)=\frac{2\|\mathbf{u}-\mathbf{p}_i\|}{c},
\end{equation}
where $c$ is the speed of light. The measured delay is modeled as
\begin{equation}
	m_i = h_i(\mathbf{u},\mathbf{p}_i) + n_i,
\end{equation}
where $n_i$ is the measurement noise. Stacking all measurements yields
\begin{equation}
	\mathbf{m}=\mathbf{h}(\mathbf{u},\mathbf{p})+\mathbf{n},
\end{equation}
where
\begin{equation}
	\mathbf{m}=[m_1,\ldots,m_N]^T,
\end{equation}
and 
\begin{equation}
	\mathbf{h}(\mathbf{u},\mathbf{p})=[h_1,\ldots,h_N]^T.
\end{equation}
\begin{rema}\label{rema2}
The measurement noise vector is modeled as $\mathbf{n} \sim \mathcal{N}(\mathbf{0}, \mathbf{Q}_n)$, where $\mathbf{Q}_n = \sigma_n^2 \mathbf{I}_N$, implying i.i.d. Gaussian noise components. Likewise, the RIS position errors are assumed to be i.i.d. Gaussian with covariance matrix $\mathbf{Q}_p = \sigma_p^2 \mathbf{I}_{DN}$, and are independent of the measurement noise \cite{zhang2025optimal}.
\end{rema}
\section{CRB Derivation and Uncertainty Analysis} \label{crb}
In this section, we show that RIS location errors do not alter the optimal RIS placement for ToF RIS-aided self-localization under the general optimality criteria of A-, D-, and E-optimality, provided that some mild conditions hold. Specifically, we prove that the CRB matrix in the presence of RIS location errors, denoted by $\mathrm{\mathbf{CRB}}$, is simply a scalar multiple of the error-free CRB matrix, denoted by $\mathrm{\mathbf{CRB}}_0$, with the scalar being independent of the system geometry. That is, $\mathrm{\mathbf{CRB}} = \alpha \hspace{2pt}\mathrm{\mathbf{CRB}}_0$, where $\alpha$ is a geometry-independent scaling factor. As a result, the optimal RIS placement remains unchanged under all three optimality criteria.

Our main objective is to estimate the source location $\mathbf{u}$, while treating the RIS positions $\mathbf{p}$
as nuisance parameters. Accordingly, we define the full unknown parameter vector as $\hm{\zeta} = [\mathbf{u}^T , \mathbf{p}^T ]^T \in \mathbb{R}^{D+DN}$. Because $\mathbf{p}$ is a nuisance random parameter, the relevant performance bound is the Hybrid CRB (HCRB), which can be written as \cite{noam2009notes}
\begin{align} \label{2041}
	\scalemath{.9}{  \mathrm{\mathbf{HCRB}} (\hm{\zeta}) =}
	\begin{bmatrix}
		\mathbf{F}_{u} \in \mathbb{R}^ {D \times D}  &  \mathbf{F}_{u, p} \in \mathbb{R}^ {D \times DN} \vspace*{3mm} \\
		\mathbf{F}_{p, u} \in \mathbb{R}^ {DN \times D}  &  \mathbf{F}_{p} \in \mathbb{R}^ {DN \times DN}
	\end{bmatrix}^{-1} ,
\end{align}
where the block matrices are given by
\begin{equation}\label{ftildu}
	\mathbf{F}_{u} = \mathbf{J}_{u}^{T} \mathbf{Q}_{n}^{-1} \mathbf{J}_{u},
\end{equation}
\begin{equation}\label{ftildu1}
		\mathbf{F}_{p} = \mathbf{J}_{p}^{T} \mathbf{Q}_{n}^{-1} \mathbf{J}_{p} + \mathbf{Q}_{p}^{-1},
\end{equation}
and
\begin{equation}\label{ftilds}
	\mathbf{F}_{u, p} = \mathbf{F}^T_{p, u} = \mathbf{J}_{u}^{T} \mathbf{Q}_{n}^{-1} \mathbf{J}_{p}.
\end{equation}
Here, $\mathbf{J}_{u}$ and $\mathbf{J}_{p}$ denote the Jacobians of the noise-free measurement $\mathbf{h}$ with respect to $\mathbf{u}$ and $\mathbf{p}$, respectively, i.e.,
\begin{equation}\label{ju}
	\mathbf{J}_{u} =\frac{\partial \hspace{1.5pt} \mathbf{h}}{\partial \hspace{1.5pt} \mathbf{u}} = \frac{2}{c} \mathbf{A}_u^T  \in \mathbb{R}^ {N \times D} ,
\end{equation}
\begin{equation}
	\mathbf{J}_{p} = \frac{\partial \hspace{1.5pt} \mathbf{h}}{\partial \hspace{1.5pt} \mathbf{p}} = -\frac{2}{c} \mathbf{A}_p^T  \in \mathbb{R}^ {N \times DN}.
\end{equation}
Moreover, 
\begin{equation}
\mathbf{A}_u =  [\mathbf{a}_1, \mathbf{a}_2, ..., \mathbf{a}_N],
\end{equation}
and 
\begin{equation}
 \mathbf{A}_p = \mathrm{BlkDiag} \left (\mathbf{a}_1, \mathbf{a}_2, ..., \mathbf{a}_N \right),
\end{equation}
where 
\begin{equation}
\mathbf{a}_i = \frac{\mathbf{u} - \mathbf{p}_i}{\|\mathbf{u}-\mathbf{p}_i\|}.
\end{equation}
Now, under the moderate RIS position error approximation in \cite{zhang2025optimal} and by applying the matrix inversion lemma (see e.g., \cite[Appendix A]{stoica2005spectral}),
the CRB for the parameter of interest $\mathbf{u}$ can be obtained from the HCRB in \eqref{2041}. In particular, the resulting bound can be expressed as
\begin{equation} \label{crbu}
	\mathrm{\mathbf{CRB}} (\mathbf{u}) = \mathrm{\mathbf{CRB}}_0 + \Delta \mathrm{\mathbf{CRB}},
\end{equation}
where $\mathrm{\mathbf{CRB}}_0 = \mathbf{F}_{u}^{-1}$ is the CRB when RIS position errors are absent, and
\begin{equation} \label{deltacrb}
	\scalemath{.95}{\Delta \mathrm{\mathbf{CRB}} =  \mathbf{F}_{u}^{-1} \mathbf{F}_{u,p}
		\left( \mathbf{F}_{p} - \mathbf{F}_{u,p}^T \mathbf{F}_{u}^{-1}  \mathbf{F}_{u,p} \right)^{-1}
		\mathbf{F}_{u,p}^T  \mathbf{F}_{u}^{-1}.}
\end{equation}
Using \eqref{ftildu} and \eqref{ju}, the error-free bound $\mathrm{\mathbf{CRB}}_0$ can be written as
\begin{equation} \label{crb0}
	\mathrm{\mathbf{CRB}}_0 = \frac{\sigma_{n}^2 \hspace{1pt}c^2}{4} \hspace{1pt}\mathbf{G}_u^{-1},
\end{equation}
where
\begin{equation}
\mathbf{G}_u = \mathbf{A}_u\mathbf{A}_u^T = \sum_{i=1}^{N} \mathbf{a}_i\mathbf{a}_i^T,
\end{equation}
is the geometry matrix collecting the directional information from the RISs to the source. For well-conditioned geometries (e.g., RISs surrounding $\mathbf{u}$ in different directions), $\mathbf{G}_u$ is positive definite. Then, invoking the Woodbury matrix identity, ${(\mathbf{D} + \mathbf{U} \mathbf{C} \mathbf{V})}^{-1} = \mathbf{D}^{-1} - \mathbf{D}^{-1} \mathbf{U} {(\mathbf{C}^{-1} + \mathbf{V} \mathbf{D}^{-1} \mathbf{U})}^{-1} \mathbf{V} \mathbf{D}^{-1}$, we can equivalently rewrite \eqref{deltacrb} as
\begin{align} \label{deltacrb2}
	\nonumber \scalemath{.95}{	\Delta \mathrm{\mathbf{CRB}} =  \mathbf{F}_{u}^{-1} \mathbf{F}_{u,p}}
	\bigg(& \scalemath{.95}{\mathbf{F}^{-1}_{p} + \mathbf{F}^{-1}_{s}  \mathbf{F}_{u,p}^T \Big( \mathbf{F}_{u} - \mathbf{F}_{u,p}\mathbf{F}^{-1}_{p}\mathbf{F}_{u,p}^T \Big)^{-1}} \\  & \times \mathbf{F}_{u,p} \mathbf{F}^{-1}_{p}  \bigg)
	\mathbf{F}_{u,p}^T   \mathbf{F}_{u}^{-1}.
\end{align}
To simplify \eqref{deltacrb2}, and following Remark~\ref{rema2}, we first express $\mathbf{F}_{u,p}$ and $\mathbf{F}_{p}$, respectively, in structured forms:
\begin{equation}\label{ftildeus2}
	\mathbf{F}_{u,p} = [\mathbf{B}_1, \mathbf{B}_2, ..., \mathbf{B}_N],
\end{equation}
\begin{equation}\label{ftildes2}
	\mathbf{F}_{p} =\mathrm{BlkDiag} \left( \mathbf{D}_1, \mathbf{D}_2, ..., \mathbf{D}_N \right), 
\end{equation}
with
\begin{equation}\label{ftildeus21}
\mathbf{B}_i =-\beta \mathbf{a}_i \mathbf{a}_i^T, 
\end{equation}
\begin{equation}\label{ftildes21}
 \mathbf{D}_i =\beta \left( \mathbf{a}_i \mathbf{a}_i^T + \frac{1}{\gamma} \mathbf{I}_D  \right), 
\end{equation}
where $\beta = \frac{4}{c^2 \sigma_{n}^2}$ and $\gamma = \frac{4\sigma_{p}^2}{c^2 \sigma_{n}^2}$. It then follows that $\mathbf{F}^{-1}_{p} = \mathrm{BlkDiag} \left( \mathbf{D}^{-1}_1, \mathbf{D}^{-1}_2, ..., \mathbf{D}^{-1}_N \right)$. Moreover, by applying the Sherman–Morrison formula to $\mathbf{D}_i$, we obtain
\begin{equation}\label{di}
	\mathbf{D}_i^{-1} = \frac{1}{\beta} \left( \gamma \mathbf{I}_D - \frac{\gamma^2 \mathbf{a}_i \mathbf{a}_i^T}{1+ \gamma} \right).
\end{equation}
Next, by introducing $\mathbf{S} = \mathbf{F}_{u,p}\mathbf{F}^{-1}_{p}\mathbf{F}_{u,p}^T$, the matrix $\Delta \mathrm{\mathbf{CRB}}$ in \eqref{deltacrb2} can be compactly written as
\begin{align} \label{deltacrb3}
	\scalemath{.93}{	\Delta \mathrm{\mathbf{CRB}} = \mathrm{\mathbf{CRB}}_0    
		\left( \mathbf{S} + \mathbf{S} \left(\mathrm{\mathbf{CRB}}_0^{-1} -\mathbf{S} \right)^{-1}  \mathbf{S}     \right)
		\mathrm{\mathbf{CRB}}_0.}
\end{align}
Using \eqref{ftildeus2}--\eqref{ftildes21}, $\mathbf{S}$ admits the decomposition
\begin{equation} \label{key1}
	\scalemath{.98}{	\mathbf{S}= \sum_{i = 1}^{N} \mathbf{B}_i \mathbf{D}^{-1}_i \mathbf{B}^T_i = \sum_{i = 1}^{N} \beta^2 \left(\mathbf{a}_i \mathbf{a}^T_i \right) \mathbf{D}^{-1}_i \left(\mathbf{a}_i \mathbf{a}^T_i \right).}
\end{equation}
Substituting \eqref{di} into \eqref{key1}, we obtain
\begin{equation} \label{key2}
	\mathbf{S} =  \beta \frac{\gamma}{1+\gamma} \sum_{i = 1}^{N} \mathbf{a}_i \mathbf{a}^T_i = \frac{\gamma}{1+\gamma} \mathrm{\mathbf{CRB}}_0^{-1}.
\end{equation}
Finally, substituting \eqref{key2} into \eqref{deltacrb3} yields $\Delta \mathrm{\mathbf{CRB}} = \gamma \hspace{2pt} \mathrm{\mathbf{CRB}}_0$, and therefore
\begin{equation} \label{crbtoafinal}
	\mathrm{\mathbf{CRB}} (\mathbf{u}) = \alpha \hspace{2pt}\mathrm{\mathbf{CRB}}_0 ,
\end{equation}
where $\alpha = 1+\gamma$. The corresponding scalar for each optimality criteria can be obtained as follows:
\begin{itemize}
	\item \textbf{A-optimality:}~~$\mathrm{tr} \lbrace \mathrm{\mathbf{CRB}} (\mathbf{u}) \rbrace = \alpha \hspace{2pt} 	\mathrm{tr} \lbrace \mathrm{\mathbf{CRB}}_0 \rbrace $,
	\item \textbf{D-optimality:}~~$\mathrm{det} \lbrace \mathrm{\mathbf{CRB}} (\mathbf{u}) \rbrace = \alpha^D \hspace{2pt} 	\mathrm{det} \lbrace \mathrm{\mathbf{CRB}}_0 \rbrace $,
	\item \textbf{E-optimality:}~~$\lambda_{\mathrm{max}} \lbrace \mathrm{\mathbf{CRB}} (\mathbf{u}) \rbrace = \alpha \hspace{2pt} \lambda_{\mathrm{max}} \lbrace \mathrm{\mathbf{CRB}}_0 \rbrace $.
\end{itemize}
Importantly, the factor $\alpha$ does not depend on the geometry (e.g., $\bar{\mathbf{p}}$). Hence, for any criterion $f(.) \in \lbrace \mathrm{tr}(.), \mathrm{det}(.), \lambda_{\mathrm{max}}(.) \rbrace$,
\begin{equation}
	\mathrm{arg min}_{\bar{\mathbf{p}}}~ f(\mathrm{\mathbf{CRB}} (\mathbf{u})) = \mathrm{arg min}_{\bar{\mathbf{p}}}~ f(\mathrm{\mathbf{CRB}}_0).
\end{equation}
Therefore, under moderate RIS position errors and the conditions in Remark~\ref{rema2}, position uncertainty induces a uniform performance loss (through a geometry-independent scaling) without changing the optimal RIS configuration.
\section{Numerical Analysis} \label{simu}
In this section, we provide numerical simulations to validate the theoretical derivations presented in Section~\ref{crb}. Specifically, we examine the scaling relation established in \eqref{crbtoafinal}, which asserts that the presence of RIS position uncertainty scales the CRB matrix by a geometry-independent factor $\alpha$. This implies that the optimal RIS placement, as determined by A-, D-, or E-optimality criteria, is invariant to the presence of RIS position errors.

To evaluate the robustness of our theoretical findings, we employ a large-scale Monte-Carlo simulation framework over a wide variety of deployment scenarios. Unlike deterministic parameter sweeps, this randomized approach allows us to evaluate the scaling law across heterogeneous localization environments and diverse RIS configurations. Although our analytical framework is general, we focus our simulations on three-dimensional scenarios, as the two-dimensional case behaves analogously. For each Monte-Carlo realization, we randomly sample the following parameters:
\begin{itemize}
	\item The number of RISs, $N \sim \mathcal{U}\{4, ..., 40\}$,
	\item The source location $\mathbf{u}$ and RIS positions $\mathbf{p}_i$, which are uniformly distributed within a sphere of radius $R=100$~m,
	\item The ToF measurement noise standard deviation, $\sigma_n$, sampled logarithmically from $[0.1~\mathrm{ns},\,10~\mathrm{ns}]$,
	\item The RIS position error standard deviation, $\sigma_p$, sampled logarithmically from $[5~\mathrm{mm},\,1~\mathrm{m}]$.
\end{itemize}
For every realization, we compute the error-free CRB matrix, $\mathrm{\mathbf{CRB}}_0 = \mathbf{F}_{u}^{-1}$, and the error-affected CRB matrix, $\mathrm{\mathbf{CRB}}$, using \eqref{crbu} and \eqref{deltacrb}. Then, the scaling relations corresponding to the A-, D-, and E-optimality criteria are numerically evaluated.

To quantify the agreement between the numerical results and the theoretical scaling law, the relative scaling error is defined as
\begin{equation} \label{relativeScalingError}
	\epsilon_f = \frac{\left|\frac{f(\mathrm{\mathbf{CRB}})}{f(\mathrm{\mathbf{CRB}}_0)}-\alpha_f\right|}{\alpha_f},
\end{equation}
where $f(\cdot)$ denotes the corresponding optimality criterion and 
\begin{equation}
	\alpha_f =
	\begin{cases}
		\alpha, & \text{A- and E-optimality},\\
		\alpha^D, & \text{D-optimality}.
	\end{cases}
\end{equation}
We perform a total of $2\times10^4$ Monte-Carlo realizations
to ensure statistical significance. Table~I reports the mean and maximum relative scaling errors associated with the three optimality criteria.

\begin{table}[t] \label{table}
	\centering
	\caption{Relative scaling errors for different optimality criteria}
	\label{tab:relative_errors}
	\begin{tabular}{|c|c|c|}
		\hline
		\textbf{Criterion} &
		\textbf{Mean Relative Error} &
		\textbf{Maximum Relative Error} \\
		\hline
		A-optimality & $4.1581\times10^{-15}$ & $2.3464\times10^{-12}$ \\
		\hline
		D-optimality & $1.0189\times10^{-14}$ & $3.0868\times10^{-12}$ \\
		\hline
		E-optimality & $5.9603\times10^{-15}$ & $2.3668\times10^{-12}$ \\
		\hline
	\end{tabular}
\end{table}

The numerical results demonstrate that the relative scaling errors remain negligibly small across all simulated deployment conditions. This verifies the scaling relation in \eqref{crbtoafinal} under the aforementioned conditions and confirms that RIS position errors uniformly scale the CRB independently of the RIS geometry. Therefore, although RIS position errors degrade localization performance, they do not alter the optimal RIS placement under A-, D-, and E-optimality criteria. A geometric illustration of the underlying reason for this performance degradation is shown in Fig.~2. The blue markers indicate the optimal nominal placement $\bar{\mathbf{p}}^{\star}$, 
which remains identical whether position errors are considered or not. 
The red markers show one realization of the actual deployed positions when i.i.d.\ Gaussian placement errors are present. The geometric mismatch between the nominal and realized locations is the
source of the performance degradation, while the optimal design 
itself is unchanged.

\begin{figure}[!b]
	\centering
	\includegraphics[width=0.45\textwidth]{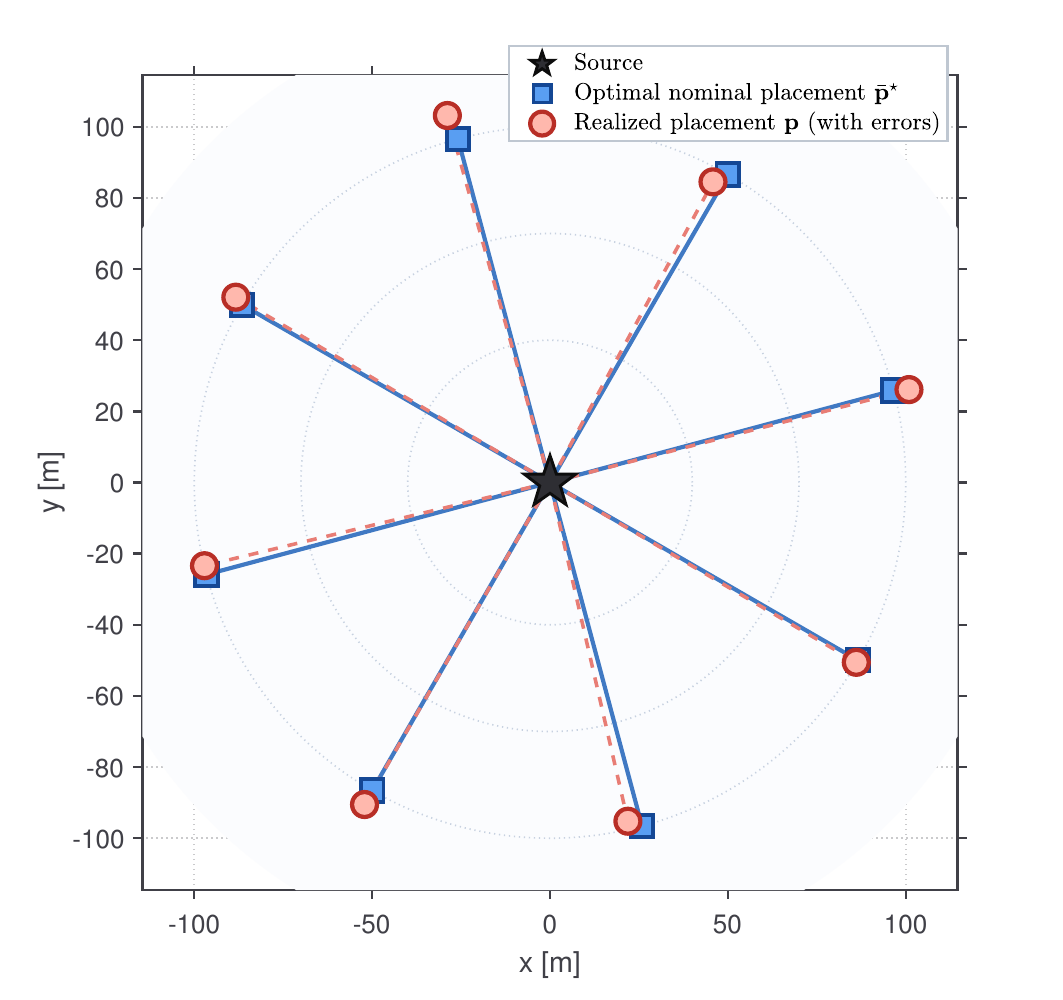}
	\caption{Geometric illustration of the performance degradation caused by RIS position errors. The blue markers show the optimal nominal positions $\bar{\mathbf{p}}^{\star}$ (identical with and without placement errors). The red markers show a typical realization of the actual deployed positions $\mathbf{p}$ in the presence of i.i.d.\ Gaussian placement errors. The source is located at the origin.}
	\label{fig:scenario}
\end{figure}

Furthermore, the results indicate that the proposed robustness property remains valid over a broad range of RIS configurations, measurement noise levels, and RIS position error levels. Hence, under moderate RIS position uncertainties, the optimal RIS deployment can be designed solely based on the CRB in the absence of RIS position errors.

\section{Conclusion}
This paper investigates the optimal placement of RISs for ToF-based self-localization in the presence of RIS deployment errors. Under the conditions of moderate RIS position errors—assuming both ToF noise and RIS position errors are Gaussian and i.i.d.—we establish a direct relationship between the source position CRB with and without RIS position uncertainty. We prove that RIS position errors scale the source position CRB by a multiplicative degradation factor that depends on the ratio of the RIS position error power to the ToF measurement noise power. Crucially, because this factor is independent of the RIS geometry, the optimal RIS placement under A-, D-, and E-optimality criteria remains identical to that in the error-free scenario.

Numerical simulations validated these theoretical results, confirming that while placement errors increase the absolute error bound by the derived factor, they do not shift the optimal coordinates for RIS deployment. This demonstrates that placement strategies designed for ideal conditions remain robust to practical deployment uncertainty.

Future work can address the impact of non-i.i.d. and non-Gaussian RIS position errors, as well as correlated ToF noise and model mismatches, including multipath and non-line-of-sight conditions. Additionally, placement robustness will be examined when ToF measurements are augmented with other observables, such as Angle-of-Arrival (AoA) or Received-Signal-Strength (RSS).
\bibliographystyle{IEEEtran}
\bibliography{myreff}

\end{document}